\documentclass[a4paper,11pt]{article}
\usepackage{pos}
\usepackage{wrapfig}

\title{The Radio detection of inclined showers at the Pierre Auger Observatory}
\ShortTitle{Auger Radio Detector}

\author*[a]{Sijbrand de Jong}
\author[b,1]{the Pierre Auger Collaboration}

\affiliation[a]{Radboud University Nijmegen and Nikhef,
  Heyendaalseweg 135, Nijmegen, The Netherlands.}
\affiliation[b]{Av. San Martin Norte 304, Malarg\"{u}e, Mendoza, Argentine.}
\note{Full Author List: \url{http://www.auger.org/archive/authors_2020_07.html}}

\emailAdd{auger\_spokespersons@fnal.gov}

\abstract{
Ultra-high-energy cosmic rays (UHECR), of energy >10 EeV, arrive at
the Earth regularly, but their sources, acceleration mechanisms,
details of 
propagation through the universe, and particle composition remain
mysteries. In addition, their interactions with the atmosphere
show an unexpectedly high muon flux compared to simulations.

To address these issues, the Pierre Auger Observatory, a hybrid
3000 km$^2$ ground based cosmic ray detector, is being upgraded, notably
adding a completely new detection layer to measure the radio
frequency emission of extensive air showers.

This Radio Detector extends the vertical shower techniques developed
in earlier radio arrays, such as the Auger Engineering Radio Array,
to horizontal showers, with a precision that is expected to be similar
to existing ground array techniques. It will provide a novel measurement
for inclined showers, complementary to the other techniques.

Details of the detection technique, the design and production of the full
1660 station Radio Detector and the expected reach in addressing
the open questions in UHECR astroparticle physics are presented.}

\FullConference{%
  40th International Conference on High Energy physics - ICHEP2020\\
  July 28 - August 6, 2020\\
  Prague, Czech Republic (virtual meeting)
}

\begin{document}
\maketitle

\section{Ultra-high-energy cosmic ray detection at the Pierre Auger Observatory}
\vspace*{-4mm}
The Pierre Auger Observatory~\cite{bib:Auger} is a 3000 km$^2$ ultra-high-energy cosmic ray (UHECR)
detector situated on the Pampa Amarilla in Argentina.
The energy range of interest is for particles with an energy of about 1 EeV and higher.
Originally Auger consisted of a Surface Detector array (SD) of 1.5 km spaced water-Cherenkov detectors
(WCD) and
a Fluorescence Detectors (FD) consisting of four stations of six fluorescence telescopes
that look for the faint light emission by showers in the atmosphere over the area covered by the SD.
Auger is now being upgraded as AugerPrime~\cite{bib:AugerPrime}
for much better particle identification on the cosmic rays.

\vspace*{-3mm}
\section{The need for particle identification of ultra-high-energy cosmic rays}
\vspace*{-4mm}
The case for more and better detection of ultra-high-energy particles from the cosmos is twofold:
First of all we see them, but we still do not know where and how they have been produced,
which must be exciting physics and astronomy.
The second reason is that we can study their interactions with the nuclei in the atmosphere at
centre-of-mass energies that are one to two orders of magnitude higher than those at the LHC.
What we are hampered by is that we do not know the particle type that is hitting us,
certainly not on an event-to-event basis.
Hence, it is impossible to track back these charged particles through the cosmic magnetic fields
to their source, based on their arrival direction,
even if we would know the cosmic magnetic fields.
It is also very difficult to interpret the interactions with unknown particles.
The particle-identification of ultra-high-energy cosmic rays is the most important reason for
the AugerPrime upgrade of the Pierre Auger Observatory.

\begin{figure}[b]
\includegraphics[width=7.5cm]{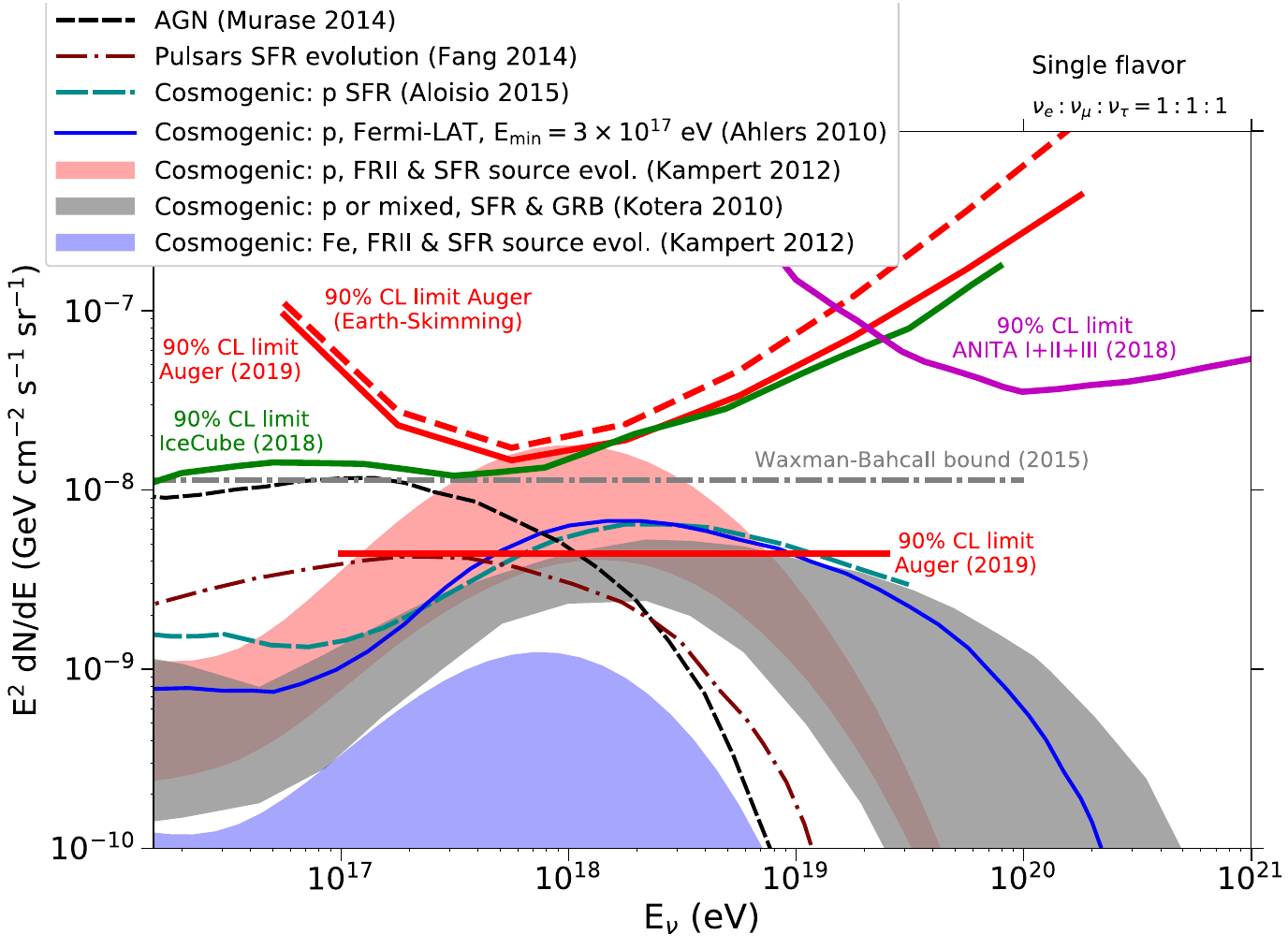}
\hfill
\includegraphics[width=7.5cm]{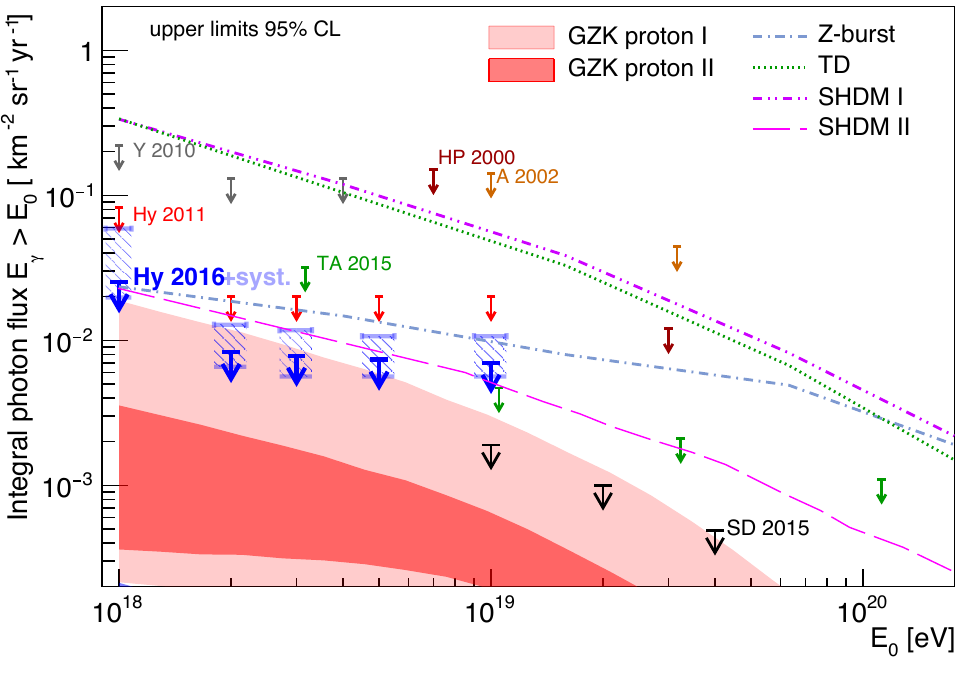}
\caption[Auger neutral limits]{\label{fig:neutrals}
                                    \em
                                    Current and future flux limits and theoretical expectations for
                                    ultra-high-energy neutrinos (left)~\cite{bib:neutrinolim}
                                    and photons (right)~\cite{bib:photonlim} as a function of energy.
                                    }
\vspace*{-10mm}
\end{figure}

Particle identification also offers opportunities for the detection of ultra-high-energy
neutrinos and photons, which have not been detected so far, but should 
exist as secondaries of
ultra-high-energy cosmic ray interactions with cosmic background photons.
The current situation for the detection of ultra-high-energy neutral particles is shown in
Fig.~\ref{fig:neutrals}.
Left are theoretical predictions for neutrino flux as a function of energy and
right for photons, together with presently observed limits, which start touching on the region of
theoretical predictions.
The aim of AugerPrime is to improve the sensitivity to ultra-high-energy neutrino and photon
detection by an order of magnitude, thereby cutting deep into the predicted range,
leading to either discovery or to serious limitations of theoretical models.

\vspace*{-3mm}
\section{Performing particle identification of ultra-high-energy cosmic rays}
\vspace*{-4mm}
There are basically two observables available to determine composition.
One is to measure the depth of maximum air shower development
into the atmosphere, $X_{\rm max}$,
which is a measure for the cross section of the incoming cosmic ray, together with a more
rapidly evolving shower for heavier cosmic rays.
The other is to determine the muon yield in the air showers, where more muons point to a
heavier cosmic ray that produced more pions in the early shower development.
A complication of the latter method is that the absolute muon yield is not yet well understood
in UHECR interactions~\cite{bib:muonproblem}, which is an interesting problem in its own right.

One way of measuring the number of muons is by adding to each SD station a layer of
Scintillator Surface Detector (SSD) with a different
response to electromagnetic particles and muons and unfold their fractions, see the contribution
of Gabriella Cataldi in these proceedings~\cite{bib:AugerPerformance}).
The other way is presented in the following:
 detection of air showers using their radio frequency radiation emission.

\vspace*{-3mm}
\section{The AugerPrime Radio Detector for ultra-high-energy cosmic rays}
\begin{wrapfigure}{r}{8cm}
\vspace*{-6mm}
\centerline{\includegraphics[width=8cm]{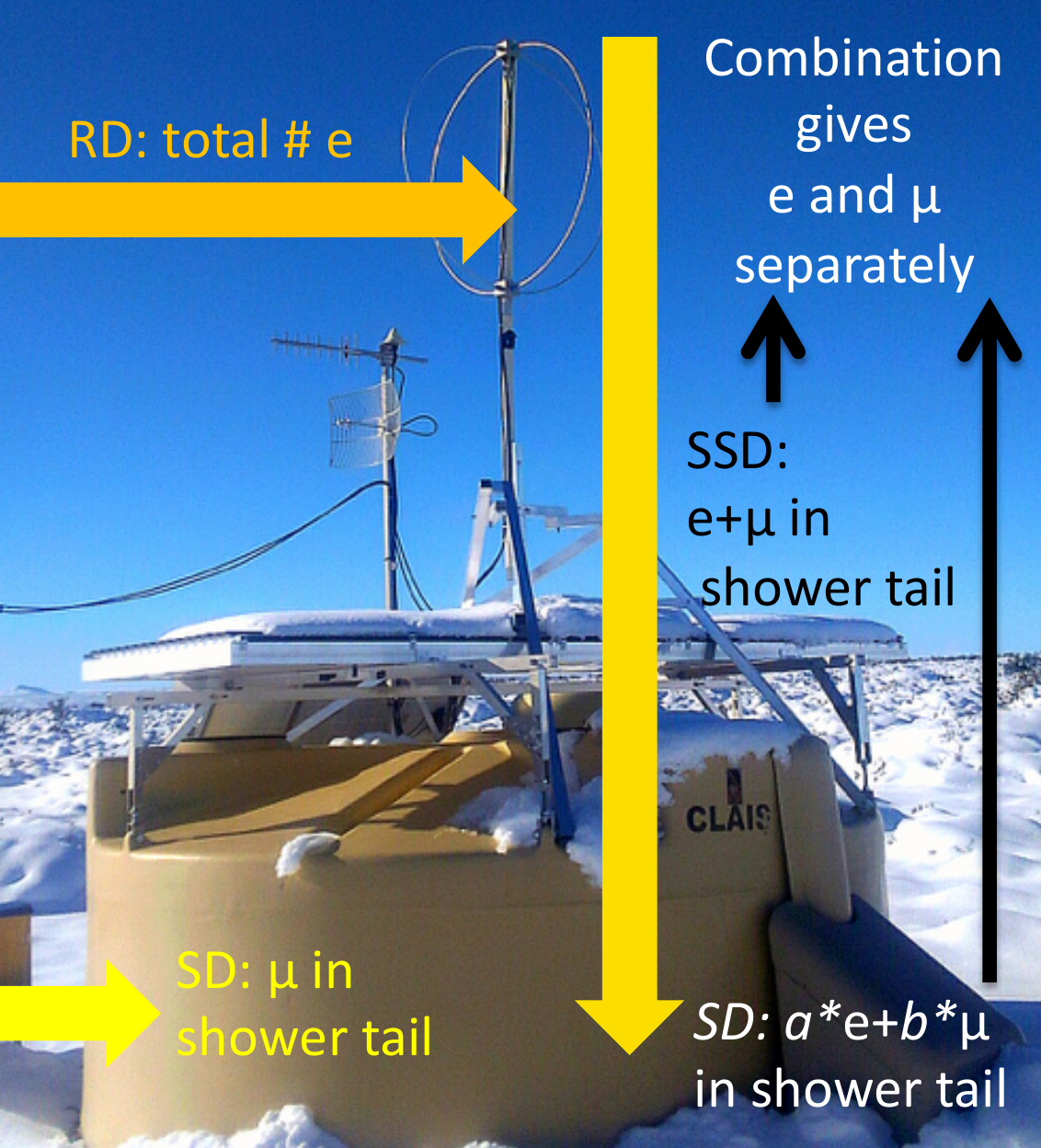}}
\caption[CR spectrum]{\label{fig:detectorstation}
                                    \em
                                    The beige SD WCD, with on top the flat metal SSD box
                                    and the two-hoop RD SALLA  on a pole.
                                    }
\vspace*{1mm}
\end{wrapfigure}

\vspace*{-4mm}
For this we will equip all 1660 Auger surface detector stations with a radio detector (RD) station,
which is a circular short aperiodic loaded loop antenna (SALLA)  for two polarisation directions on top,
sensitive in the frequency range of 30--80~MHz.
Their analogue signal will be processed and digitised and digitally processed by its
own electronics board.
The processed data will then be merged into the existing, yet upgraded,
readout and data acquisition system of the SD.
The power for the radio detector will be taken from the existing SD station power system,
where the solar panels will be replaced by more powerful ones, an operation that was
anyway necessary to meet the power requirements of the upgraded SD electronics.
The full setup of a SD, SSD and RD station is shown in Fig.~\ref{fig:detectorstation}.

Because the radio frequency pulse emitted by air showers travels unimpeded through the
atmosphere, the RD will be able to measure the electromagnetic component of inclined showers,
where the shower has already been fully absorbed before reaching the ground.
Particle identification is possible by comparing the muon yield as measured by the SD station
with the electromagnetic shower signal in the RD. This is schematically illustrated by the
horizontal arrows on the left in Fig.~\ref{fig:detectorstation}.
This method is complementary to the muon--EM shower components deconvolution that
can be done for vertical showers with the SD and SSD combination, as illustrated by the
vertical arrow in the middle of the picture.

The two methods complement each other for different inclinations, opening the field of view,
and thereby also the acquired

\noindent
number of events as a function of declination \\[-5mm]
\noindent
\begin{wrapfigure}{r}{8cm}
\vspace*{-10mm}
\centerline{\includegraphics[width=8cm]{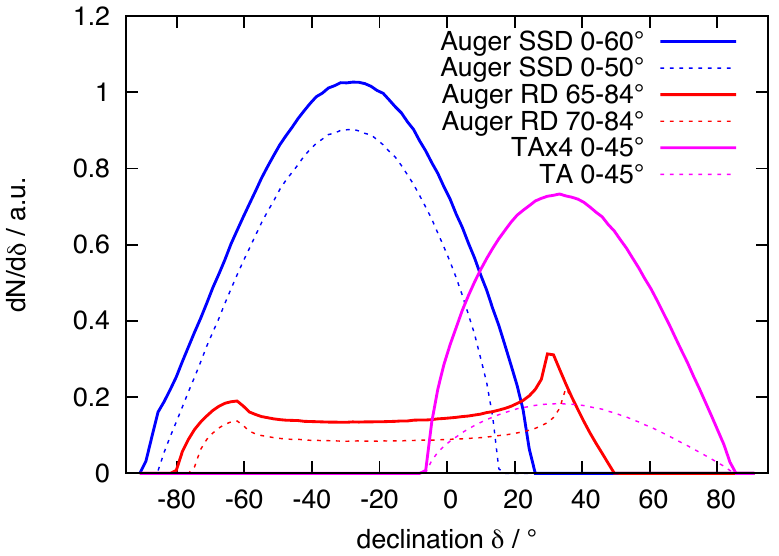}}
\caption[CR spectrum]{\label{fig:eventsbydecl}
                                    \em
                                    Expected relative number of measured events as a function of
                                    declination for the SSD plus SD combination and for the
                                    RD plus SD combination.
                                    }
\vspace*{-4mm}
\end{wrapfigure}
statistics,
as can be seen by the expected
in Fig.~\ref{fig:eventsbydecl}.
The additional number of events at UHE will be many tens of thousands,
which is an impressive number in this field.

\vspace*{-3mm}
\section{AERA experience}
\vspace*{-4mm}
Our confidence to deploy this large radio detector array is based in 20 years of preparatory work, notably in the past decade with the Auger Engineering Radio Array, AERA.
While the full radio detector will cover all 1660 existing surface detector station, AERA was a 10\%
test bed of standalone radio detector stations with different inter-station spacings,
different antenna designs, and different types of readout and DAQ electronics.

From AERA
and other radio detection test set-ups we learned in quantitative detail what the emission
mechanisms of the radio frequency radiation is and how they interfere~\cite{bib:RadioOverview}.
We are now able to measure energy, arrival direction and composition of
cosmic rays to a precision that is better than that of the existing surface detector and nearly matches
that of the existing fluorescence detector that can only be operated in requires moonless, cloudless nights
for less than 15\% of the time.
The radio detector can be operated 24/7.
One of the notable observations is that the footprint of the measurable radio signal for
vertical air showers is of the order of 100-200 m in diameter.

\begin{figure}[b]
\vspace*{-3mm}
\includegraphics[width=7cm]{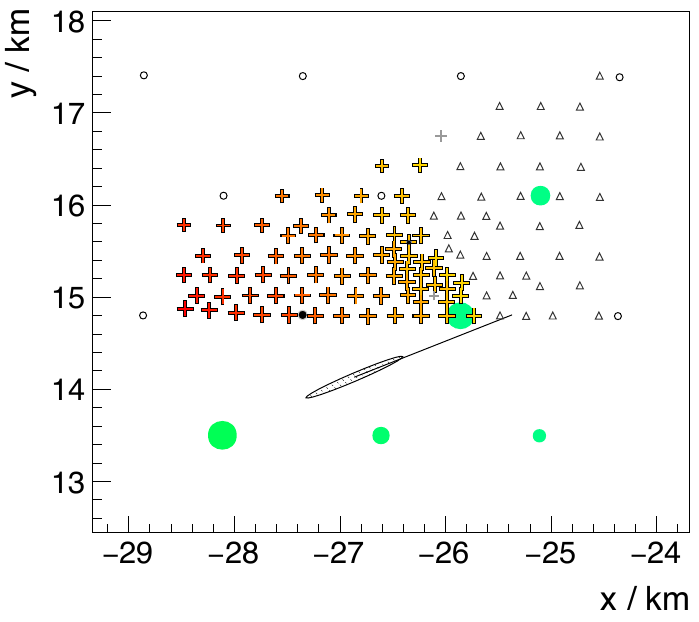}\hfill
\includegraphics[width=7cm]{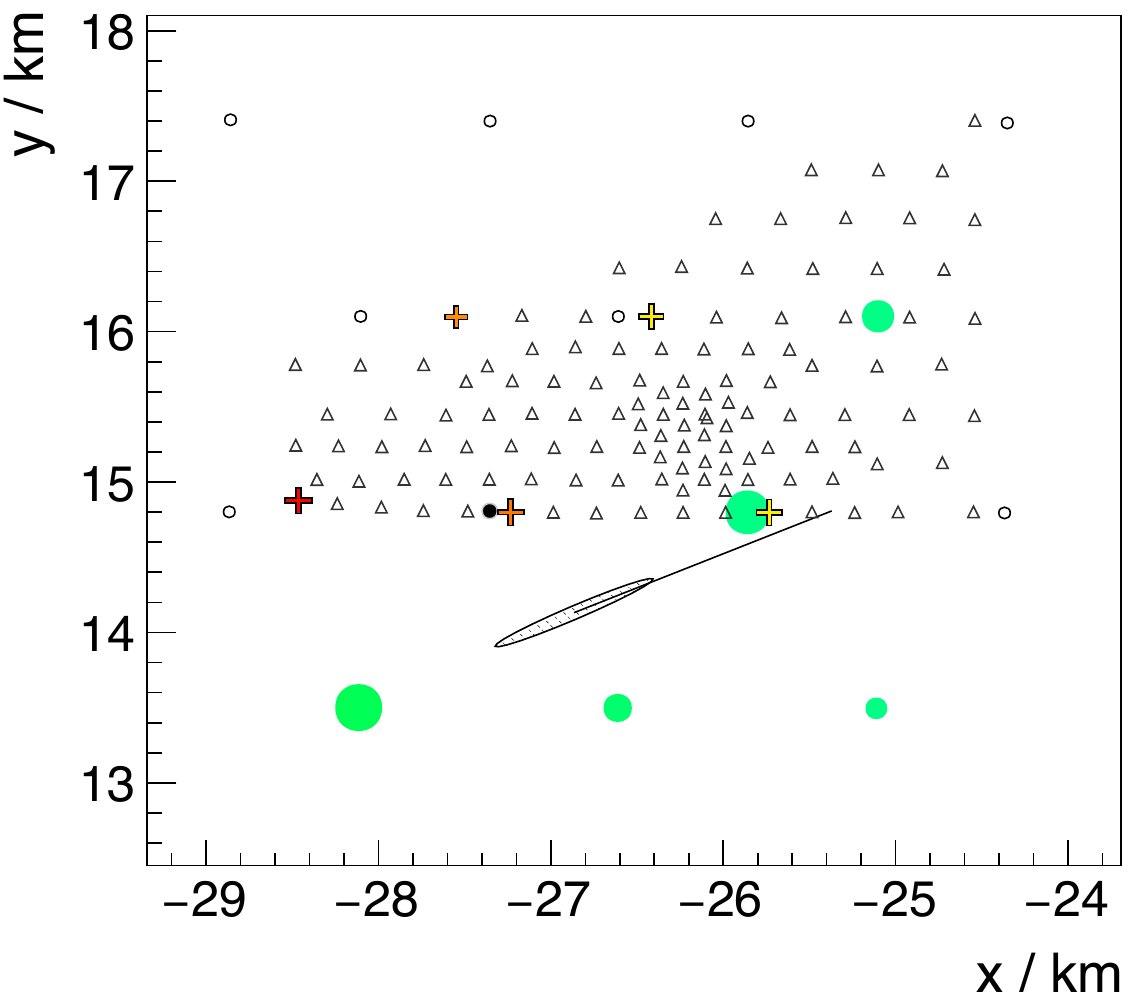}\\[-8mm]
\caption[CR spectrum]{\label{fig:AERAinclined}
                                    \em
                                    On the left, an inclined shower as detected in AERA.\cite{bib:AERAhas}
                                    The green spots give the SD station
                                    (spot size corresponds to energy) and the crosses give the
                                    radio detector measurements.
                                    On the right, only the hit radio detector stations in a 1.5~km grid, like for the
                                    AugerPrime RD, are shown as red crosses.~
                                    }
\vspace*{-5mm}
\end{figure}

While the small footprint requires a densely spaced array for vertical showers,
for an inclined shower, with zenith angle above 50$^{\circ}$-60$^{\circ}$,
the footprint extends over many kilometres, as illustrated by a typical event
 in Fig.~\ref{fig:AERAinclined}.
The footprint of the inclined showers is well matched to the 1.5 km water-Cherenkov
 surface detector station spacing in Auger.
This footprint and first approaches to analysing these inclined shower have been tested with AERA data,
where we have  already more than 500 inclined shower events recorded.
In Fig.~\ref{fig:AERAinclined} one of these events is shown, with the crosses indicating
 the many radio detector stations
with a signal above threshold.
When imposing a 1.5 km grid still 5 stations would be hit, even though part of the shower is outside
of the fiducial area. Hence, for a shower contained in a larger detector such as being realised
with the RD about 10 or more station will be hit, giving excellent reconstruction prospects.

\vspace*{-3mm}
\section{Progress in design and prototyping of the AugerPrime RD}
\vspace*{-4mm}
\begin{wrapfigure}{r}{10cm}
\vspace*{-4mm}
\includegraphics[height=4cm]{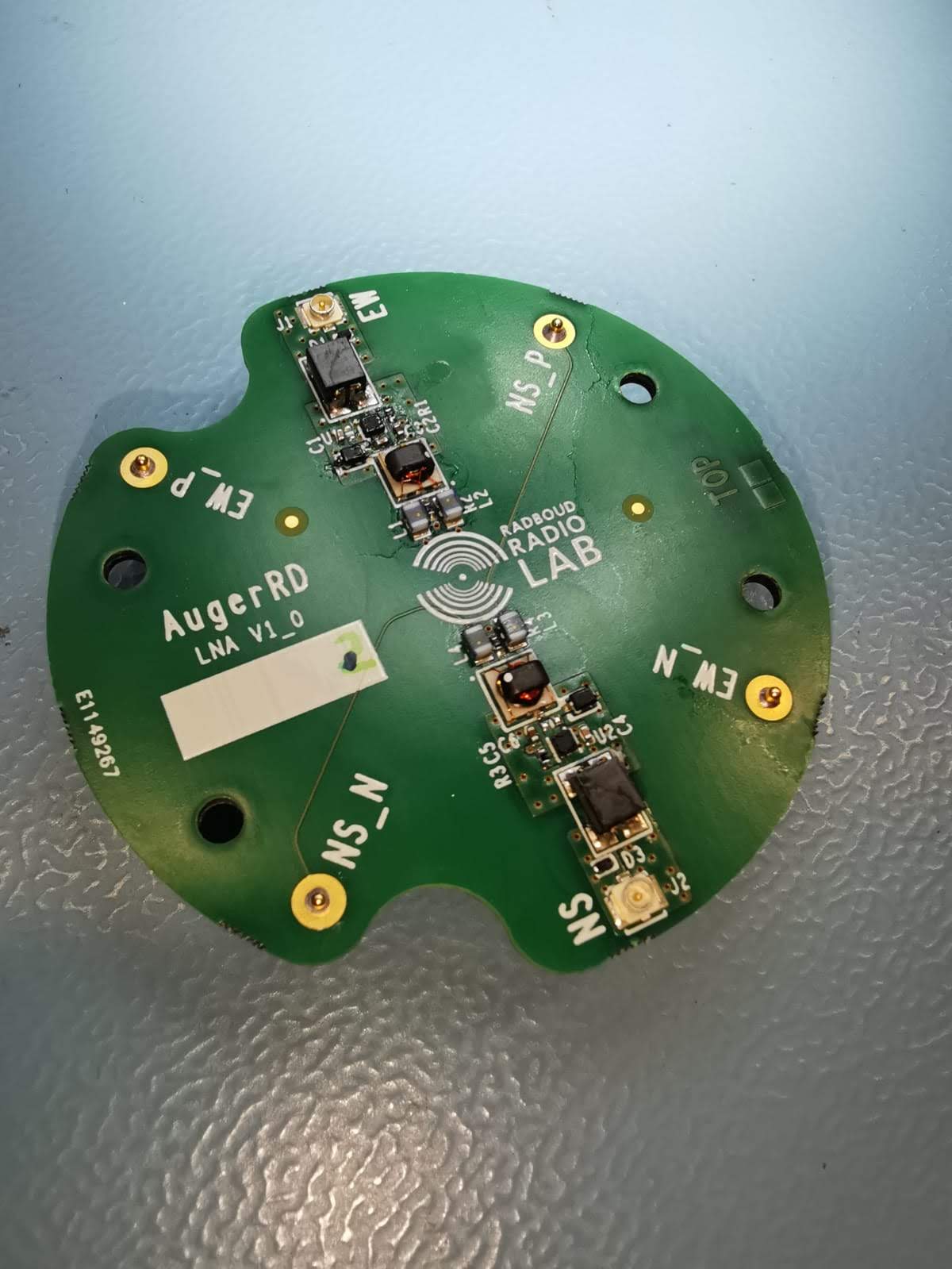}
\includegraphics[height=4cm]{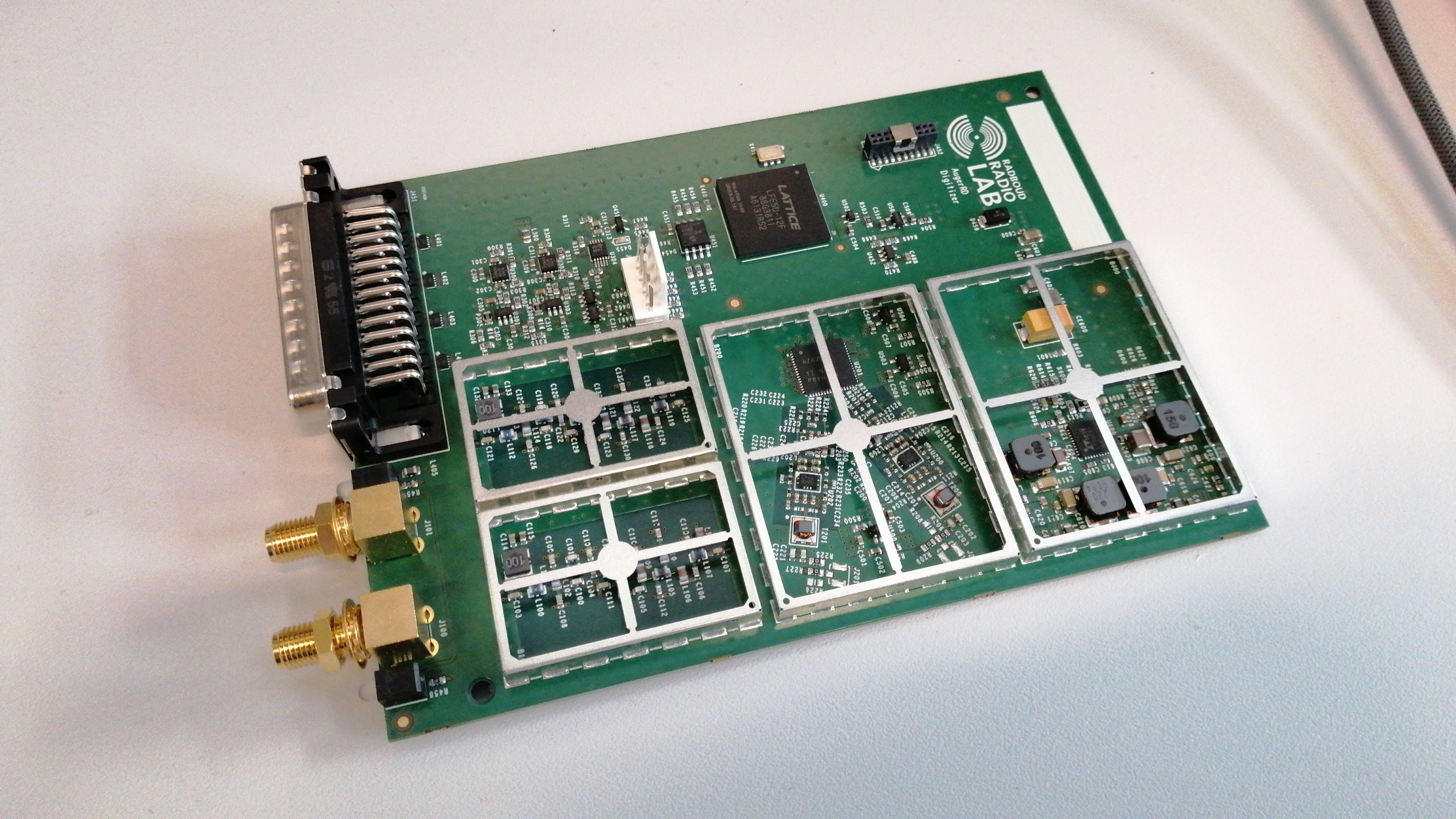}
\caption[CR spectrum]{\label{fig:electronics}
                                    \em
                                    Low noise front-end amplifier (LNA) board to be mounted
                                    on the antenna on
                                    the left and the digitiser board housed in the SD station electronics
                                    dome on the right.
                                    }
\end{wrapfigure}
At the moment the mechanical design of the radio stations is complete and has been
extensively tested in the lab and in the field.
The electronics design also nears completion, where the last work is done on
optimising the gain and noise performance to be able to make absolute signal calibrations
on the radio emission background of the Milky Way. Prototypes for the
 low noise front-end amplifier (LNA) and the
digitiser board are shown in Fig.~\ref{fig:electronics}.
As can be seen in the picture of the digitiser board, special care has been taken to
shield from electromagnetic interference, both from outside the board as well as
limiting the emission from the board itself to a minimum.
For optimal performance of the RD, also the the upgraded electronics for the SD and
SSD have been carefully optimised for minimal electromagnetic interference
emission in the relevant frequency bands.
The reconstruction of the RD event data will be part of the standard AugerPrime
event reconstruction. The necessary programming code for this presently being developed
and tested. Examples of the preliminary event reconstruction are shown in Fig.~\ref{fig:RTevents}.

\vspace*{-3mm}
\section{Prototype experience in the field}
\vspace*{-4mm}
There have been 10 prototypes in the field since November 2019
(one of them depicted in Fig.~\ref{fig:detectorstation}),
of which 7 are being read out, where we have obtained very valuable information
to optimise the design and experience for deployment.

\begin{figure}[t]
\vspace*{-6mm}
\includegraphics[width=7.4cm]{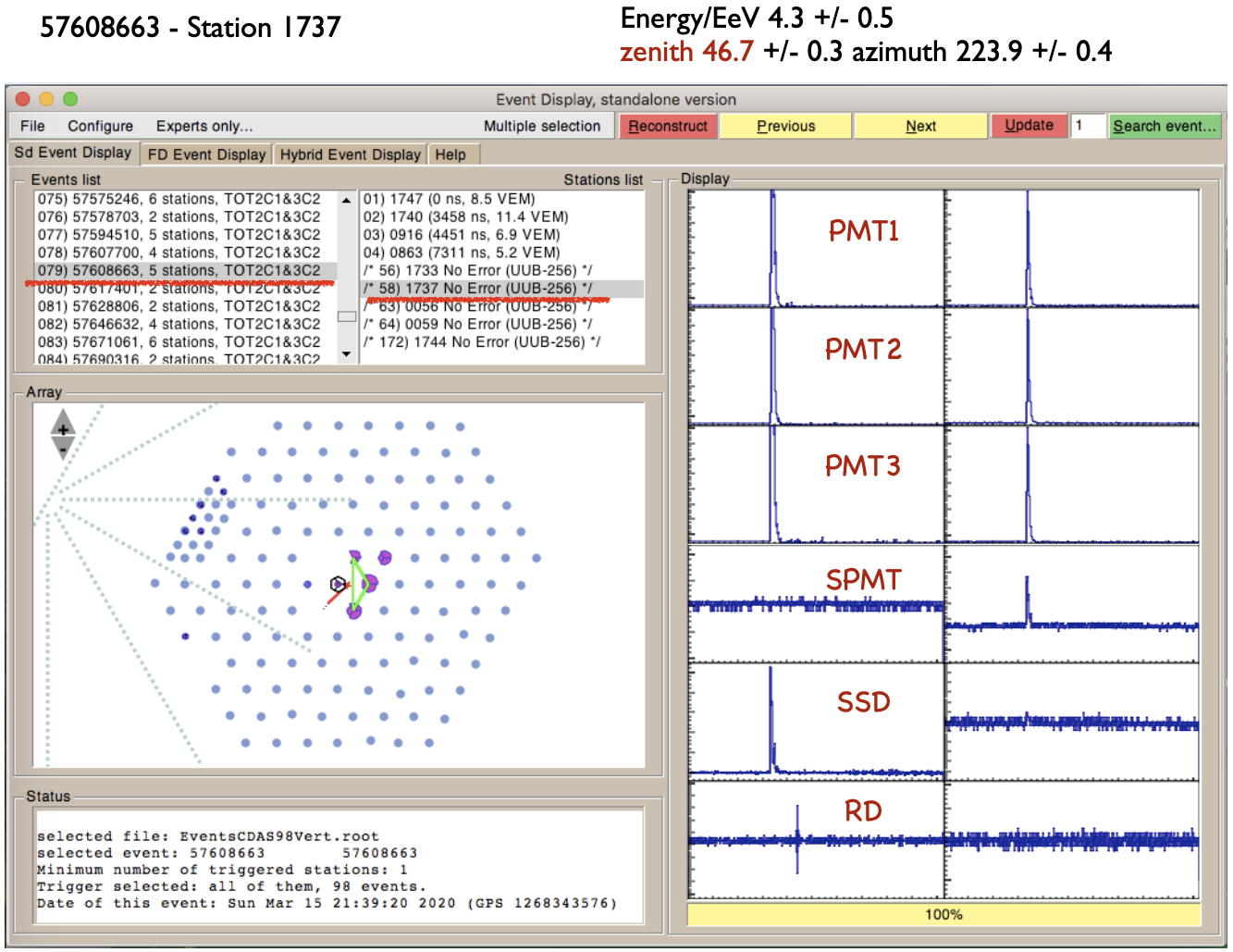}\hfill
\includegraphics[width=7.4cm]{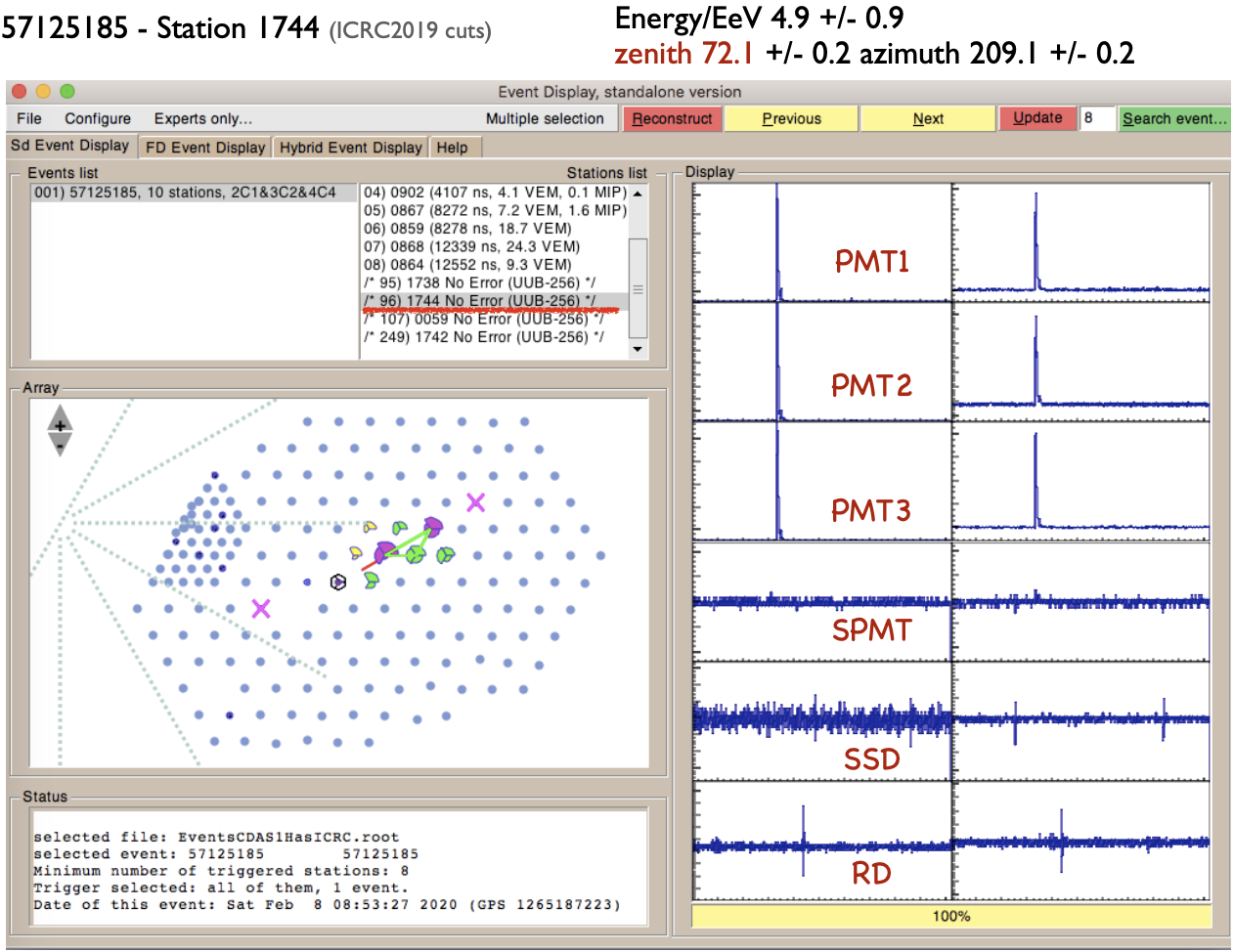}
\caption[CR spectrum]{\label{fig:RTevents}
                                    \em
                                    Event displays of preliminary reconstructions of
                                    a vertical (left) and a horizontal (right) shower
                                    as recorded in the AugerPrime RD prototype engineering array,
                                    together with SD and SSD.
                                    }
\vspace*{-3mm}
\end{figure}

In Fig.~\ref{fig:RTevents} event displays are shown for events recorded jointly by the
SD, SSD and RD.
Both events have an energy of about 4-5 EeV.
The left one has a zenith angle of 47$^{\circ}$ and the right one 72$^{\circ}$.
The grid on the left hand side shows which SD and RD stations have a signal above threshold.
On the right hand side the signal time traces for one of the SD station equipped with an SSD module
and a RD antenna are plotted.
The three SD station photomultipliers are indicated by PMT1, PMT2 and PMT3 and the coincident signal
peaks indicate significant light generation in the WCD.
A small photomultiplier, which will also be installed as part of the AugerPrime upgrade to deal with
the very large SD WCD signals near the shower core, has not yet been installed
in this station.
The SSD module time trace is also indicated and nicely in coincidence with the WCD
signal.
The bipolar RD signal is also clearly visible, albeit with a time offset that has meanwhile been
understood and corrected for.
The more vertical event in the left event display of Fig.~\ref{fig:RTevents} has, in addition to the SD
WCD signals, detections in the SSD module and RD antenna. This event is in the declination
overlap region of the SSD and RD and will allow for cross calibration and combination of the
information from these detectors.
The more horizontal event on the right hand side of Fig.~\ref{fig:RTevents} has no signal in the SSD
module because of the limited cross section of the 1~cm thick scintillator projected on the shower
plane. The RD antenna signal in this case provides the additional information that will allow to
measure the muon yield, which is most of the SD WCD signal in this case, with respect to the
electromagnetic shower content as measured by the RD. 

\vspace*{-3mm}
\section{Conclusion and outlook}
\vspace*{-4mm}
The production of the radio detectors is expected to start by the end of this year or the beginning of 2021.
A few months delay has been caused by the Covid-19 pandemic. The deployment of the AugerPrime
RD will be in 2021 and 2022, with a decade or more to collect data afterwards.

\end{document}